\documentclass[conference,9th]{IEEEtran}
\IEEEoverridecommandlockouts
\usepackage{cite}
\usepackage{amsmath,amssymb,amsfonts}
\usepackage{algorithmic}
\usepackage{graphicx}
\usepackage{textcomp}
\usepackage{xcolor}
\usepackage{physics}
\usepackage{setspace}
\usepackage{cite}
\usepackage[ruled,vlined]{algorithm2e}
\usepackage{tikz}
\def\checkmark{\tikz\fill[scale=0.4](0,.35) -- (.25,0) -- (1,.7) -- (.25,.15) -- cycle;} 

\usepackage{orcidlink}
\usepackage{tabularx}
\usepackage{enumitem}
\usepackage{adjustbox}
\usepackage{graphicx}
\usepackage{amsmath}
\usepackage[font=small,skip=0pt]{caption}
\usepackage{comment}

\usepackage{booktabs} 
\usepackage{url}
\def\BibTeX{{\rm B\kern-.05em{\sc i\kern-.025em b}\kern-.08em
    T\kern-.1667em\lower.7ex\hbox{E}\kern-.125emX}}

\makeatletter
\renewcommand\footnotesize{\fontsize{7.6pt}{8.6pt}\selectfont}
\makeatother

\begin{document}

\title{QFNN-FFD: Quantum Federated Neural Network for Financial Fraud Detection
}

 \author{\IEEEauthorblockN{Nouhaila Innan
 \textsuperscript{1,2}, Alberto Marchisio
 \textsuperscript{1,2}, Mohamed Bennai
 \textsuperscript{3}, and Muhammad Shafique
 \textsuperscript{1,2}
 } \IEEEauthorblockA{\textsuperscript{1}eBRAIN Lab, Division of Engineering, New York University Abu Dhabi (NYUAD), Abu Dhabi, UAE\\ \textsuperscript{2}Center for Quantum and Topological Systems (CQTS), NYUAD Research Institute, NYUAD, Abu Dhabi, UAE\\ 
 \textsuperscript{3}Quantum Physics and Spintronic Team, LPMC, Faculty of Sciences Ben M'sick,\\ Hassan II University of Casablanca, Morocco\\ nouhaila.innan@nyu.edu, alberto.marchisio@nyu.edu, mohamed.bennai@univh2c.ma, muhammad.shafique@nyu.edu\\ }}

\maketitle

\begin{abstract}
This work introduces a Quantum Federated Neural Network for Financial Fraud Detection (QFNN-FFD), an advanced framework merging Quantum Machine Learning (QML) and quantum computing with Federated Learning (FL) for financial fraud detection. Using quantum technologies' computational power and the robust data privacy protections offered by FL, QFNN-FFD emerges as a secure and efficient method for identifying fraudulent transactions within the financial sector. Implementing a dual-phase training model across distributed clients enhances data integrity and enables superior performance metrics, achieving precision rates consistently above 95\%. Additionally, QFNN-FFD demonstrates exceptional resilience by maintaining an impressive 80\% accuracy, highlighting its robustness and readiness for real-world applications. This combination of high performance, security, and robustness against noise positions QFNN-FFD as a transformative advancement in financial technology solutions and establishes it as a new approach for privacy-focused fraud detection systems. This framework facilitates the broader adoption of secure, quantum-enhanced financial services and inspires future innovations that could use QML to tackle complex challenges in other areas requiring high confidentiality and accuracy.

\end{abstract}

\begin{IEEEkeywords}
Quantum Neural Network, Quantum Federated Learning, Quantum Machine Learning, Fraud Detection, Finance, Classification
\end{IEEEkeywords}

\section{Introduction}
In the rapidly evolving financial technology landscape, privacy is a fundamental pillar, crucial for upholding the trust and integrity of financial transactions and services \cite{peng2021privacy}. As digital transactions become more prevalent, the volume of sensitive data handled by financial institutions grows exponentially, making robust privacy measures indispensable \cite{chand2024credit}. The emergence of Quantum Machine Learning (QML) marks a transformative era \cite{qml,schuld2015introduction,huang2021power,Innan_Grover_2024,N.Innan,innan2023enhancing,dutta2024qadqn,innan2025optimizing,innan2025qnn}, promising computational capabilities by exploiting quantum physics~\cite{zaman2023survey}, while simultaneously raising pivotal concerns about privacy and data security. 
This paper introduces the Quantum Federated Neural Network for Financial Fraud Detection (QFNN-FFD), a framework that integrates the quantum-enhanced processing power of Quantum Computing (QC) with the privacy-preserving attributes of Federated Learning (FL). The synergy of QML with FL jointly improves the efficiency and accuracy of detecting fraudulent activities, while safeguarding sensitive financial data against the ever-looming threats of breaches and unauthorized access.

\begin{figure}[htpb]
    \centering
\includegraphics[width=1\linewidth]{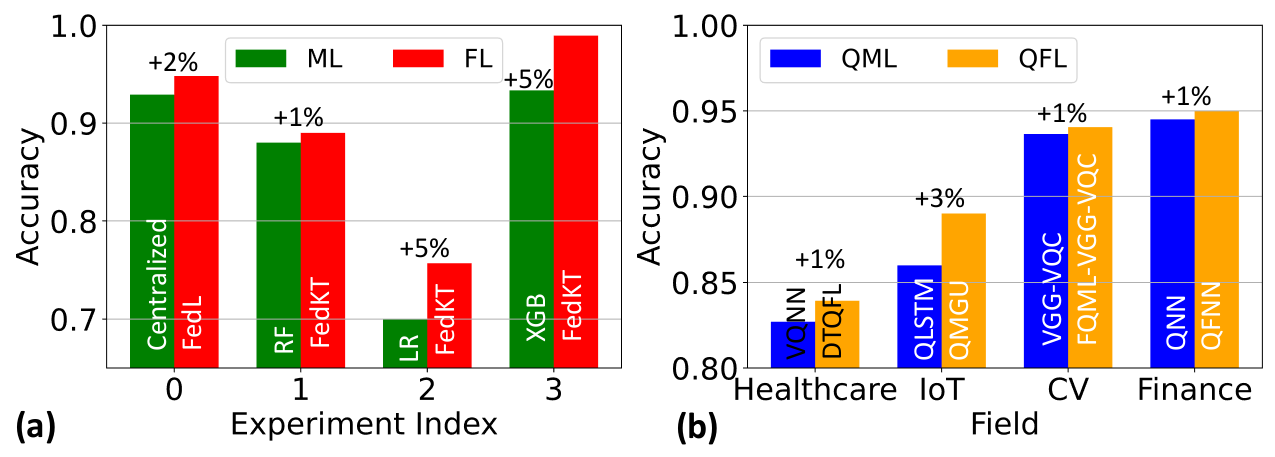}
    \caption{ 
    Comparison of ML and FL accuracies in classical and QC contexts across various fields and experiments. Panel \textbf{(a)} illustrates the performance of different experiments within the finance sector. Panel \textbf{(b)} compares QML with QFL across four domains: healthcare, IoT, computer vision, and finance. In classical computing contexts, FL generally demonstrates superior performance compared to ML \cite{shingi2020federated,wang2024novel}. In QC contexts, QFL exhibits slight improvements over QML \cite{qu2023dtqfl, qu2024qfsm, chen2021federated}. These findings highlight the potential of QFL and provide a compelling rationale for its adoption, particularly in the finance sector.}

    \label{compa}
\end{figure}

QFNN-FFD demonstrates the potential of quantum technologies in addressing real-world economic challenges and sets a new benchmark for privacy-centric approaches in the fintech domain. By deploying this framework, financial institutions can potentially employ the advantages of QC—such as the potential rapid processing of large datasets—while also benefiting from the decentralized nature of FL, which keeps sensitive data localized and reduces the risk of central points of failure. As shown in Fig. \ref{compa}, Quantum Federated Learning (QFL) has shown superior performance in various fields \cite{zhang2022federated,yun2022slimmable,zhao2023non,chu2023cryptoqfl}, prompting our decision to implement it in finance. Our framework has demonstrated its capability to enhance both accuracy and privacy protection through comparative analysis with existing models~\cite{ref1, ref2, kyriienko2022unsupervised}. This approach meets and often surpasses current industry standards, providing a scalable, secure framework that adapts seamlessly to diverse operational environments while maintaining high accuracy in fraud detection under various conditions.
\begin{figure}[htpb]
    \centering
    \includegraphics[width=1\linewidth]{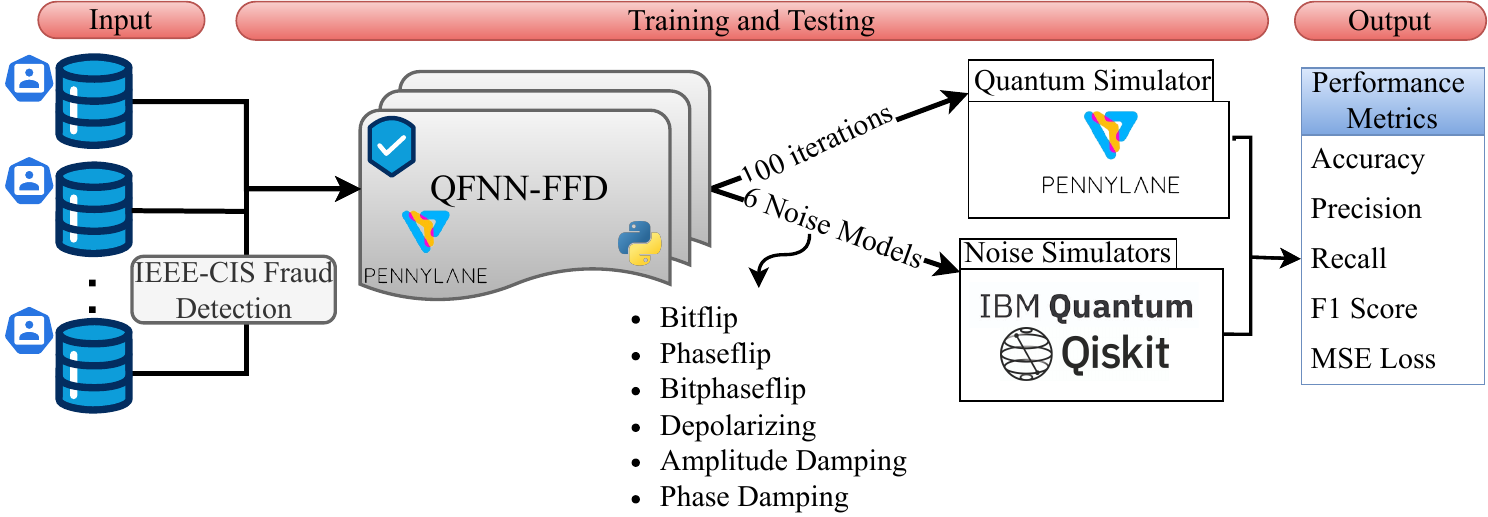}
    \vspace{0.01cm}
    \caption{ 
    The QFNN-FFD process flow. The diagram outlines the end-to-end workflow from input through to output. Datasets are processed and fed into the QFNN-FFD, built upon the PennyLane library. The model undergoes training and testing for 100 iterations, incorporating a variety of noise models using noise simulators from IBM's Qiskit. The quantum simulator within PennyLane is utilized to emulate a quantum environment. The output is evaluated based on performance metrics, including accuracy, precision, recall, F1 score, and mean squared error loss, providing a comprehensive assessment of the model's capability to detect fraudulent transactions.}
    \label{workflow}
\end{figure}

\textbf{
Our contributions significantly impact the fintech sector by providing a secure framework that adapts to various operational environments while maintaining high accuracy in fraud detection under different conditions, and can be listed as follows and shown in Fig. \ref{workflow}:
}
\begin{itemize}[leftmargin=*]

\item Introducing a novel QFNN-FFD that uniquely combines QML algorithms with FL architecture to enhance both the computational capabilities and the privacy aspects of fraud detection systems, ensuring that sensitive financial data remains within its local environment.
\item Demonstrating superior analytical capabilities by analyzing complex transaction patterns more effectively than traditional models, comparative experimental results reveal that QFNN-FFD consistently outperforms existing fraud detection systems in terms of accuracy, thereby establishing a new benchmark for the industry.
\item Recognizing the challenges posed by quantum decoherence and noise by testing our QFNN-FFD across six different quantum noise models to validate its robustness ensures that our framework is not only theoretically but also practically viable in real-world QC environments, maintaining high performance under various noise conditions.
\end{itemize}

\section{Background and Related Works}

FL is a Machine Learning (ML) paradigm in which multiple parties \cite{pmlr-v54-mcmahan17a, konevcny2016federated,zhang2021survey}, termed clients, collaborate under the oversight of a central server to address an ML task without exchanging their raw data. 

\begin{figure}[h]
    \centering
    \includegraphics[width=1\linewidth]{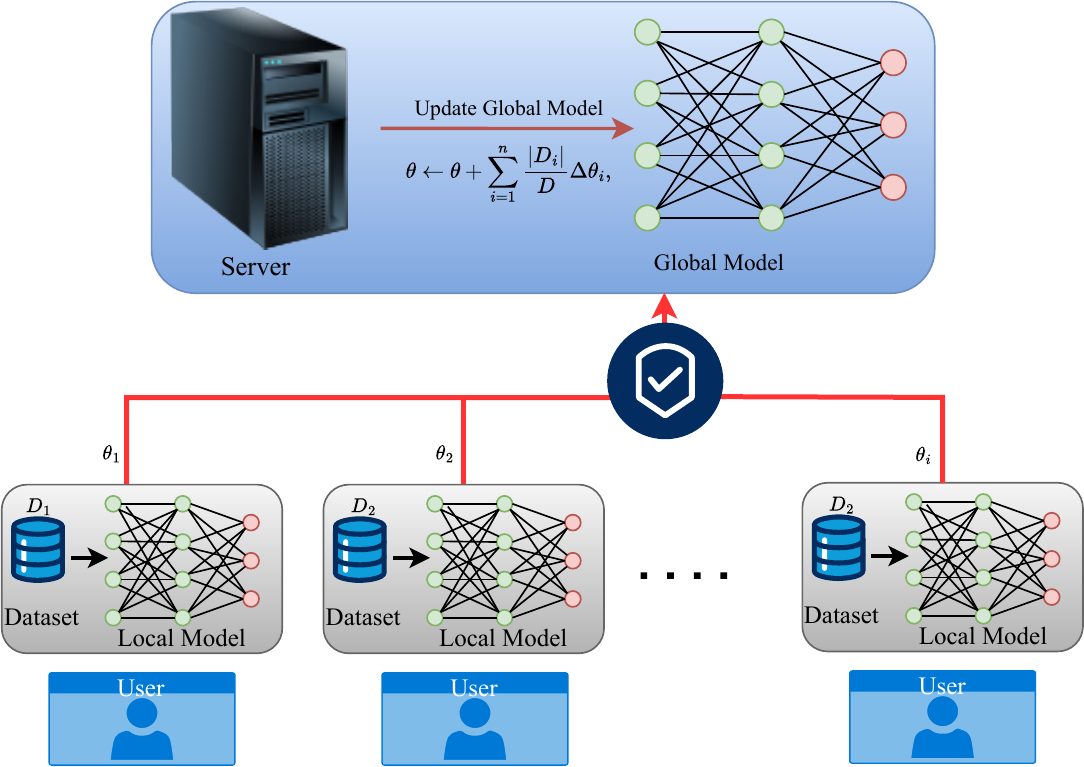}
    \vspace{0.05cm}
    \caption{ 
    Schematic representation of the FL architecture. The diagram shows multiple users (clients), each with their local dataset, independently training local models. These models are then transmitted as model updates to a central server. The server aggregates these updates to improve the global model, which is then distributed back to the users for further refinement. This cycle ensures data privacy and security, as raw data never leaves the local premises of each user.}
    \label{FL}
\end{figure}

As illustrated in Fig. \ref{FL}, clients contribute model updates, computed from their local datasets, to the server. Mathematically, each client $i$ computes an update $\Delta \theta_i$ based on its local data $D_i$:
\begin{equation}
\Delta \theta_i = -\eta \nabla L(\theta; D_i),
\end{equation}
where $\eta$ is the learning rate and $L$ is the loss function evaluated with the ML model parameters $\theta$. These updates $\Delta \theta_i$ are then sent to the central server, which aggregates them to update the global model using a weighted average:
\begin{equation}
\theta \leftarrow \theta + \sum_{i=1}^n \frac{|D_i|}{D} \Delta \theta_i,
\end{equation}
where $D = \sum_{i=1}^n |D_i|$ represents the total size of data across all clients, and $|D_i|$ is the size of the local dataset of client $i$. This aggregation method effectively mitigates concerns related to privacy, data security, and data access rights, which are particularly pertinent when dealing with sensitive information scattered across disparate locations.

The progression of FL into the QC domain has precipitated the inception of QFL \cite{chen2021federated,Chehimi2023foundations,chen2024introduction}. This methodology exploits quantum mechanics' distinctive properties to augment privacy and computational efficiency. In \cite{chehimi2022quantum}, the study delineated the first fully operational QFL framework capable of processing exclusively quantum data. This innovation indicated the establishment of the inaugural quantum federated dataset, facilitating the collaborative learning of quantum circuit parameters by quantum clients in a decentralized manner—a cornerstone in adapting quantum technologies to federated contexts.

Subsequently, the notion of dynamic QFL frameworks was advanced in \cite{yun2022slimmable}, which introduced the Slimmable QFL. This framework was designed to adapt to varying network conditions and constraints on computing resources by dynamically modulating the training parameters of Quantum Neural Networks (QNNs).
The research outlined in \cite{li2021quantum} proposed a quantum protocol that used the computational capacities of remote quantum servers while safeguarding the privacy of the underlying data. 

It is essential to recognize the expansive applications of QFL across various industries and how these applications introduce specialized implementations in sectors requiring high data privacy and computational precision. Particularly in the financial industry, where the confidentiality and integrity of data are paramount, the transition from general data protection to targeted fraud detection represents a critical evolution of QFL capabilities. 

The effectiveness of QFL in securely managing and processing data within healthcare and genomics, as explored in \cite{innan2024fedqnn}, serves as a foundation for its application in the more complex and sensitive realm of financial transactions. This broad applicability underscores the potential of QFL to enhance privacy and computational efficiency in highly effective scenarios.

Advancing into financial fraud, significant research has been conducted to apply QC and QML in detecting financial fraud. In \cite{kyriienko2022unsupervised}, they developed quantum protocols for anomaly detection, applying them to credit card fraud. 
Furthermore, in \cite{grossi2022mixed}, they explored using a Quantum Support Vector Machine (QSVM) for real-world financial data, presenting one of the first end-to-end applications of QML in the financial sector. 
As the application of QML in fraud detection advances, several innovative approaches have emerged. For instance, in \cite{innan2024financial,alami2024comparative}, they explored using QML models, including the Variational Quantum Classifier (VQC) and different QNNs. These models showed promising results in classifying fraud and non-fraud transactions, demonstrating QML's potential in financial applications. 
In \cite{wang2022integrating}, the study addressed the latency in traditional fraud detection systems by implementing a QML approach using a Support Vector Machine (SVM) enhanced with quantum annealing solvers. 
In \cite{mitra2021experiments}, they discussed a hybrid model that combines QNNs with classical neural networks to enhance fraud detection capabilities. 

Generative adversarial networks (GANs) have also been adapted to quantum settings to tackle the instability and inefficiency of classical sampling methods. \cite{herr2021anomaly} introduced variational quantum-classical Wasserstein GANs (WGANs), which incorporated a hybrid quantum-classical generator with a classical discriminator. This model was effective on a credit card fraud dataset, providing competitive performance with classical counterparts in terms of F1 score.
Further advancing the field, in \cite{pena2022fraud}, they presented an approach using data re-uploading techniques to train single-qubit classifiers that perform comparably to classical models under similar training conditions. 
Moreover, in \cite{priyaradhikadevi2023credit} and \cite{icaart24}, they highlighted the real-time challenges in fraud detection.

These studies collectively demonstrate the growing capability of QML to enhance fraud detection but often neglect the aspect of data privacy in their computational frameworks. Most QML models focus primarily on computational advantages without integrating robust privacy measures. Our QFNN-FFD framework addresses this gap by combining the privacy-preserving features of FL with the power of QC. By ensuring that data remains local and only aggregate updates are shared, our framework enhances the security and privacy of the distributed learning process, setting a new standard in applying quantum technologies to sensitive financial operations.
\section{QFNN-FFD Framework Design}
In this section, we introduce a novel QFNN-FFD framework that integrates the quantum computational capabilities of QML with the distributed, privacy-preserving nature of FL, as described in Algorithm~\ref{alg:QFNN}.
\begin{algorithm}[h]
\caption{QFNN-FFD Framework}\label{alg:QFNN}
\begin{small}
\KwData{QNN circuit, dataset split among $N$ clients, learning rate $\eta=0.1$, maximum local iterations $T$.}
\KwResult{Accuracy, precision, recall, F1 score, and loss}
Initialization: Parameters $\theta$ randomly initialized in $[0, 1]$\;
\For{each client $i=1$ to $N$}{
 Initialize local model parameters $\theta_i \leftarrow \theta$\;
 \For{each local iteration $t=1$ to $T$}{
  \For{each batch in local dataset}{
   Encode data into quantum states\;
   Apply QNN circuit with current parameters $\theta_i$\;
   Perform quantum measurements to obtain classical outputs\;
   Calculate loss using MSE\;
   Optimize $\theta_i$ using Adam optimizer with learning rate $\eta$\;
  }
  Evaluate local model on validation set and adjust $\theta_i$\;
  If convergence criteria are met, exit loop early\;
 }
 Synchronize and send optimized local parameters $\theta_i$ to central server\;
}

On central server:\;
Aggregate local parameters to update global model;
Broadcast updated global parameters $\theta$ back to each client\;

\For{each client $i=1$ to $N$}{
  Update local model parameters $\theta_i \leftarrow \theta$\;
}

Evaluate model performance on a global validation set to ensure generalization\;
\end{small}
\label{alg}
\end{algorithm}

\subsection{QNN Circuit Design and QFL Integration}
\begin{figure}[htpb]
    \centering
    \includegraphics[width=1\linewidth]{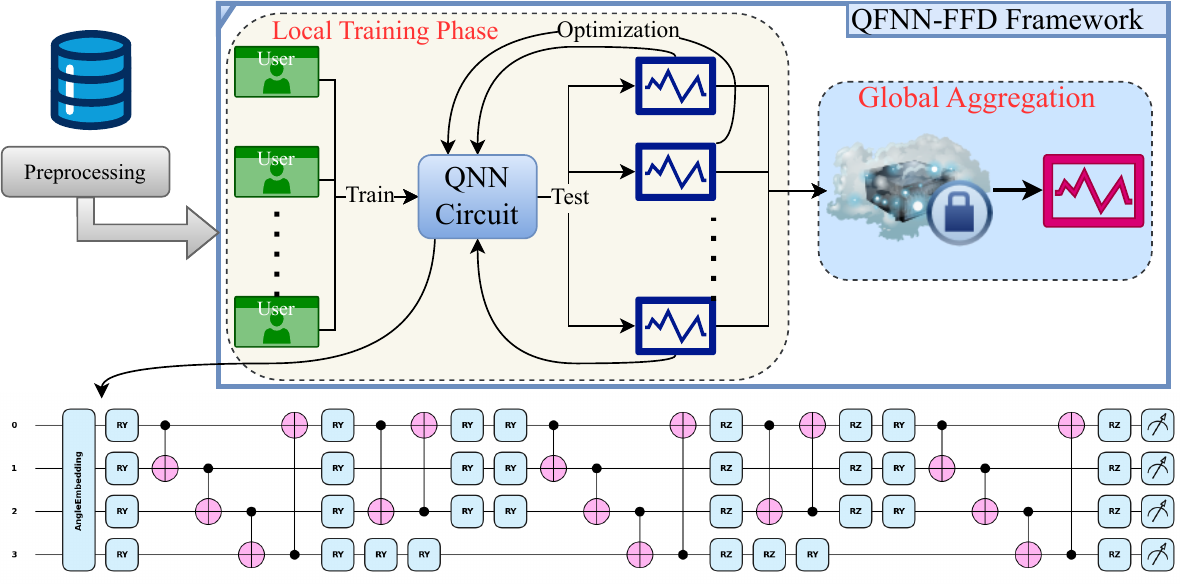}
    \caption{
    An overview of the QFNN-FFD framework. This flowchart presents the multi-stage process, beginning with data preprocessing and distribution to various users. Each user independently conducts a local training phase on a QNN circuit, followed by an optimization stage. The optimized local models are then transmitted to a central cloud server for global aggregation, culminating in an enhanced federated model. The lower part of the figure illustrates the quantum circuit's structure, showcasing the intricate interplay of qubits and quantum gates (rotations and CNOT gates) during the computation process.}
    \label{framework}
    \vspace{-10pt}
\end{figure}

\begin{figure}[htpb]
    \centering
    \includegraphics[width=1\linewidth]{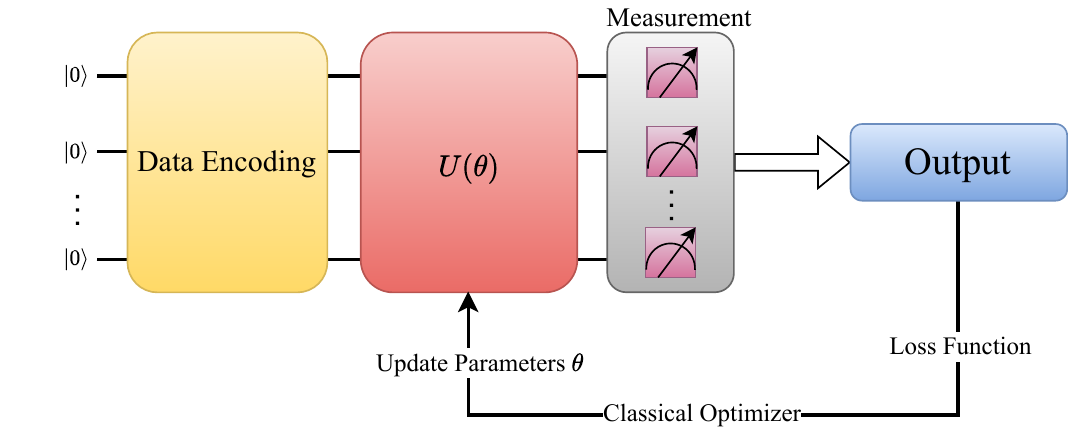}
    
    \caption{ 
    General schematic of a QML model workflow. The process begins with qubits in the zero state $(|0\rangle)$. The qubits undergo data encoding to represent the input data in quantum states. Then, a parametrized quantum circuit, $U(\theta)$, transforms the qubit states, where $\theta$ represents tunable parameters. The transformed quantum states are measured, converting quantum information into classical output. This output is evaluated using a predefined loss function, and a classical optimization algorithm iteratively adjusts $\theta$ to minimize the loss, thereby refining the QML model's performance.}
    \label{qml}
    \vspace{-10pt}
\end{figure}

Central to this approach is a QNN circuit, shown in Fig. \ref{framework}. The QNN model has demonstrated its powerful capabilities in various applications, particularly fraud detection. Like typical QML models, as shown in Fig. \ref{qml}, it begins with data encoding, followed by a sequence of quantum operations that form the core of the processing circuit, and concludes with measurement to extract actionable insights \cite{schuld2014quest, kashif2024hqnet, beer2020training, abbas2021power}. 

The QFNN-FFD framework operates on data distributed across $N$ clients, with each client possessing an Identically and Independently Distributed (IID) subset of the overall dataset. This uniform distribution ensures that all clients train the models under similar data conditions, which prevents the need for central data aggregation and enhances data privacy.

Training of the QFNN-FFD is directed in a federated manner, where local models on each client are independently trained using their data subsets.

In the local model, the first step is to encode classical data into quantum states through angle encoding. Each data feature $x_{i,j}$ from the vector $\mathbf{x}_i$ for client $i$ is mapped onto two rotation angles, $\theta_{i,j}$ for the $R_y$ rotation and $\phi_{i,j}$ for the $R_z$ rotation. These rotations are then applied to the qubits sequentially to modify both their phase and orientation:
\begin{equation}
R(\theta_{i,j}, \phi_{i,j}) = R_y(\theta_{i,j}) R_z(\phi_{i,j}),
\end{equation}
where $R_y(\theta_{i,j}) = e^{-i \theta_{i,j} Y/2}$ and $R_z(\phi_{i,j}) = e^{-i \phi_{i,j} Z/2}$, with $Y$ and $Z$ representing the Pauli-Y and Pauli-Z matrices, respectively.

We apply a series of controlled operations to achieve an entangled quantum state that captures correlations between different features. One effective method is using a sequence of CNOT gates, which create entanglements between successive qubits:
\begin{equation}
    U_{\text{ent}} = \prod_{k=1}^{n-1} \text{CNOT}_{k,k+1},
\end{equation}
where $\text{CNOT}_{k,k+1}$ applies a CNOT gate between the $k$-th and $(k+1)$-th qubits. This sequence creates a chain of entanglements across the qubit register, which is crucial for leveraging quantum correlations.

This setup ensures that the quantum states are intricately linked, which is crucial for capturing complex correlations in the dataset. The full quantum state preparation for client $i$ is thus represented by:
\begin{equation}
\ket{\psi_i} = \left(\bigotimes_{j=1}^{n} R_y(\theta_{i,j}) R_z(\phi_{i,j}) \right) \cdot \text{CNOT} \ket{0}^{\otimes n}.
\end{equation}

\subsection{Optimization and Training Process}

The Adam optimizer is integral to the training process of our QFNN-FFD framework due to its adaptive learning rate capabilities, which significantly enhance convergence speed and efficiency. The Adam optimizer's update rule is particularly well-suited for the demands of quantum circuit training and is defined as follows:
\begin{equation}
    \theta_{t+1} = \theta_t - \frac{\eta}{\sqrt{\hat{v}_t} + \epsilon} \hat{m}_t,
\end{equation}
where $\eta$ represents the learning rate, $\hat{m}_t$ and $\hat{v}_t$ are the estimates of the first and second moments of the gradients, respectively, and $\epsilon$ is a small constant to avoid division by zero. This configuration allows each parameter update to be adjusted dynamically based on the individual gradients' variability, providing a tailored approach to parameter optimization.

For gradient computation in the QNN circuit, we employ the parameter-shift rule \cite{wierichs2022general}. This method is particularly well-suited for QML models because it provides an exact estimation of the gradient, avoiding the issues associated with finite differences and stochastic gradient methods. 
The gradient of the QNN circuit parameter $\theta_i$ is given by: \begin{equation} \frac{\partial L}{\partial \theta_i} = \frac{L(\theta_i + \frac{\pi}{2}) - L(\theta_i - \frac{\pi}{2})}{2}, \end{equation} where $L(\theta)$ denotes the loss function evaluated at the parameter-shifted values. 

In the context of our QFNN-FFD, the Adam optimizer's role extends to effectively minimizing the Mean Squared Error (MSE) loss function during the training process. The MSE loss function is crucial for calibrating the model's predictive accuracy and is expressed as:
\begin{equation}
L(\theta) = \frac{1}{m} \sum_{j=1}^{m} (y_j - \hat{y}_j(\theta))^2,
\end{equation}
where $m$ is the batch size, $y_j$ are the actual labels of transactions, and $\hat{y}_j(\theta)$ represents the predicted labels output by the model. This loss function quantifies the error between the model's predictions and the true labels, guiding the optimizer to focus on reducing these discrepancies.
The optimization process iterates through a maximum of $T$ local iterations, refining the model's ability to discern fraudulent transactions accurately. 
\subsection{Parameter Aggregation and Model Evaluation}
Following local optimization, each client's parameters $\theta_i$ are transmitted to a central server. They are aggregated through a simple averaging process to update the global model parameters $\theta$. This cyclic process of local optimization and global aggregation iteratively enhances the QFNN-FFD's performance, which is evaluated on a global validation set for generalizability and efficacy.
The mathematical foundation of parameter optimization within the QFNN-FFD employs the Adam optimizer, adjusting $\theta$ as 
\begin{equation}
\theta_{t+1} = \theta_t - \eta \cdot \text{Adam}(\nabla_{\theta} L(\theta_t))    
\end{equation}
where $\text{Adam}(\nabla_{\theta} L(\theta_t))$ calculates the adjustment based on the gradient of the loss function with respect to the parameters $\theta$ at iteration $t$. This optimization ensures a gradual refinement of the model's parameters. 

After the local training phases, the optimized parameters $\theta_i$ from each client are securely aggregated at a central server using a federated averaging algorithm:
\begin{equation}
\theta_{\text{global}} = \frac{1}{N} \sum_{i=1}^N \theta_{i},
\end{equation}
This aggregation step effectively combines insights from all the distributed models, enhancing the global model's generalizability and robustness, steering the QFNN-FFD towards higher accuracy in fraud detection (see Algorithm \ref{alg}).

The globally updated parameters are redistributed to all clients for further training, cycling through local optimization, and global aggregation to progressively improve the QFNN-FFD's performance. This iterative process enhances computational efficiency and maintains strict privacy standards.

Integrating QML with FL in our QFNN-FFD framework fosters the high-efficiency processing of complex financial data and upholds stringent data privacy standards. This dual advantage, coupled with the model's mathematical rigor and strategic parameter optimization, positions the QFNN-FFD as an effective tool in the fight against financial fraud, marking substantial progress in applying QC to real-world challenges in the financial sector.
\subsection{Computational Complexity Analysis}
The computational complexity of our framework arises from both the depth of the QNN circuit and the FL process. Our QNN implementation processes each data point through multiple layers of single-qubit rotations and entanglement operations. This results in a per-sample computational complexity of $\mathcal{O}(nL)$, where $n$ is the number of qubits and $L$ is the total number of layers. Gradient computation using the parameter-shift rule requires two additional circuit evaluations per parameter, leading to a per-iteration complexity of $\mathcal{O}(P)$, where $P$ is the total number of parameters and is proportional to $nL$. Considering $T$ local iterations over a dataset of size $D$ per client, the overall computational complexity per client becomes $\mathcal{O}(TDP)$. The federated averaging process adds minimal overhead, with a communication complexity of $\mathcal{O}(P)$ per client per round. Despite the intensive quantum computations, the workload is effectively distributed across clients, and efficient gradient computation ensures the framework remains computationally feasible for practical applications in FFD.
\section{Results and Discussion}
\subsection{Experimental Setup}
In our study, we utilize the IEEE-CIS Fraud Detection dataset \cite{data}. It is divided into two primary files: identity and transaction, linked by TransactionID. It encompasses both numerical and categorical features essential for identifying fraudulent activities. The preprocessing steps begin with optimizing the dataset's memory usage by refining data types, significantly reducing its memory footprint. This is followed by a detailed analysis of missing values, which helps identify and quantify missing data to inform our approach to managing these instances. Subsequently, features are categorized and processed: categorical variables undergo one-hot encoding, while numerical variables are standardized. To counteract the class imbalance between fraud and non-fraud instances, an up-sampling technique is employed to ensure equitable representation of both classes \cite{chawla2002smote}.

Our QFNN-FFD is implemented using PennyLane for model architecture and Qiskit for simulating quantum noise \cite{pennylane,qiskit2024}, enabling a realistic QC environment. We conduct extensive hyperparameter tuning, exploring various configurations before selecting the optimal settings. The framework, structured around four qubits, employs the Adam optimizer ($\eta$=0.1) across dual training phases—local and global—with up to 100 iterations for each across 15 clients. This setup is characterized by 32 initially random parameters, which are optimized through evaluations on a training set comprising 115,386 instances (80\% of the total dataset of 144,233 instances) and a validation set comprising 28,847 instances, which is 20\% of the total dataset. We focus on binary classification accuracy and MSE as key metrics. Operational deployment occurs within 
an environment characterized by a configuration consisting of 4 virtual CPUs (vCPUs), 25 gigabytes (GB) of RAM, and a single NVIDIA Tesla V100 virtual GPU (vGPU). This setup offers a balance of processing power, memory capacity, and advanced GPU acceleration—crucial factors for efficiently handling the intensive computations required by our QFNN-FFD framework.
\subsection{Accuracy and Loss Analysis}

\begin{figure}
    \centering
    \includegraphics[width=1\linewidth]{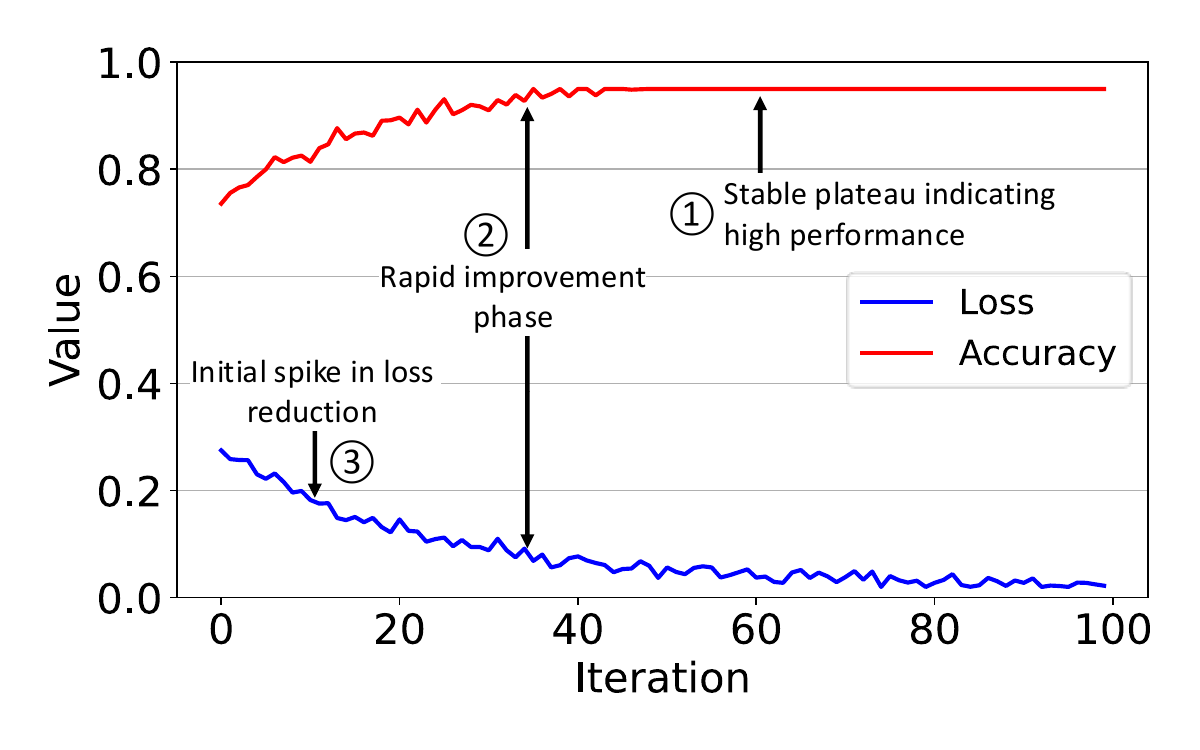}
        \caption{ 
    Evolution of validation metrics as a function of iteration count. The plot illustrates the optimization trajectory over 100 iterations, with validation accuracy demonstrating an upward trend towards convergence, and loss exhibiting a reciprocal decrease, indicative of the model's improving generalization on unseen data.}
    \label{res}
    \vspace{-10pt}
\end{figure}

The validation accuracy and loss trends for the QFNN-FFD, as shown in Fig.~\ref{res}, provide valuable insights into the model's performance over iterations, which is the average outcome of 10 trials of QFNN, which ensures the reliability of the results by accounting for variability in the model's performance. 
Initially, the model's accuracy begins at 0.735 and demonstrates a steady upward trend, culminating in a plateau of 0.95 at \textcircled{\raisebox{-0.9pt}1}, consistently maintained from iteration 35 onwards. This performance plateau signifies that the framework not only swiftly attains a high confidence level in its fraud detection capabilities but also sustains this efficacy over time. Alongside, the validation loss diminishes from an initial 0.275 to 0.02, reflecting the model's enhanced precision in identifying fraudulent transactions. 

This reduction in validation loss is significant as it suggests a substantial enhancement in the model's ability to differentiate between fraudulent and legitimate transactions with minimal error, thereby reducing the likelihood of costly false positives. The pronounced improvement in accuracy and reduction in loss observed between iterations 10 and 20 at \textcircled{\raisebox{-0.9pt}2} marks a critical learning phase for the model. By iteration 35, the model achieves and upholds a state of high accuracy and minimal loss at \textcircled{\raisebox{-0.9pt}3}, indicative of its robust learning mechanism and stability. This phase showcases the effective convergence of the quantum and FL components, optimizing the model's parameters for high-stakes decision-making environments. The sustained model performance beyond the 35th iteration underscores the QFNN-FFD's ability for dependable and steady fraud prediction within QC environments.

Moreover, the robust validation performance of QFNN-FFD highlights its practical applicability. The high validation accuracy suggests effective pattern recognition is crucial for fraud detection, while the low and stable loss indicates minimized rates of false positives and negatives—essential for the operational deployment of any fraud detection system. This balance is particularly important in financial contexts where the cost of false negatives can be extraordinarily high. Given the observed performance plateau, implementing an early exit strategy in training could economize on computational resources without compromising effectiveness, optimizing overall efficiency. This strategy underscores the framework's capability to deliver high performance while efficiently managing computational demands, setting a new standard for privacy-focused, quantum-enhanced financial services.

\subsection{Local Client Validation Accuracy Analysis}
\begin{figure}[htpb]
    \centering
    \vspace{-15pt}
    \includegraphics[width=1\linewidth]{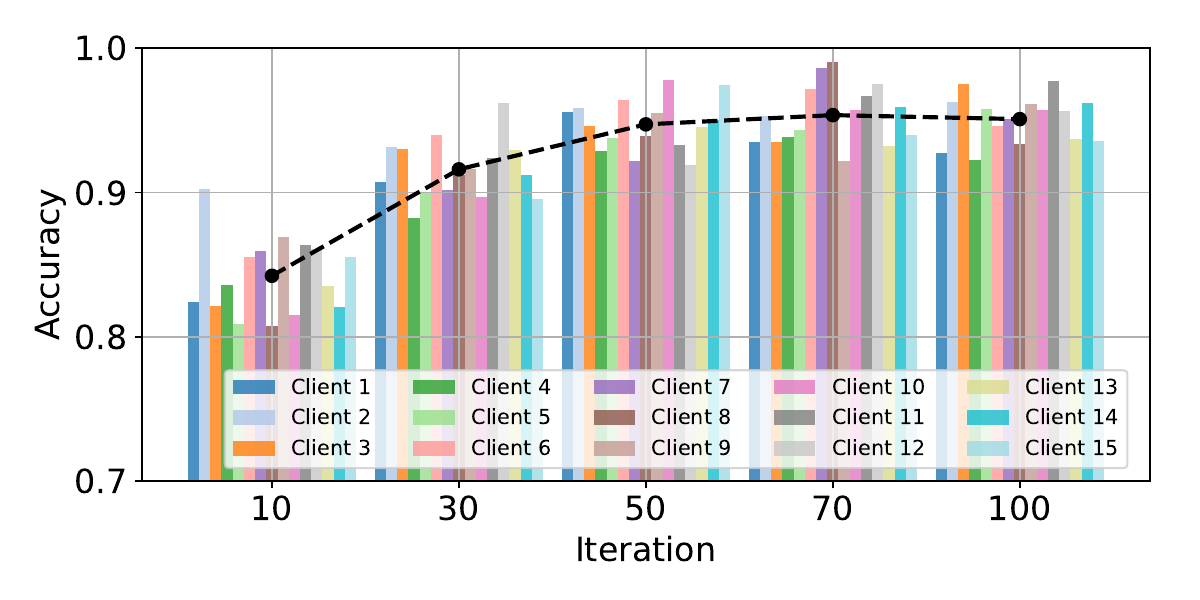}
    \caption{
    Validation accuracy trends for individual clients across selected iterations (10, 30, 50, 70, 100). Each bar represents the accuracy of a specific client at a given iteration, with the dashed black line indicating the mean accuracy trend across all clients. The figure illustrates the initial variability in client performance, followed by a convergence toward higher accuracy levels as training progresses, with most clients stabilizing around or above 0.9 accuracy by the 100th iteration. Minor fluctuations observed in some clients highlight the influence of data heterogeneity and model sensitivity on individual learning outcomes.}
    \label{local}
    \vspace{-10pt}
\end{figure}
The validation accuracy trends across individual clients reveal important aspects of our framework's performance on distributed datasets (batches). As shown in Fig. \ref{local}, the initial variability in accuracies suggests that each client's model interacts differently with its local data, influenced by the unique characteristics of the data and the stochastic elements in the training process, such as random initialization and the adaptive optimizer's behavior.
As training progresses, the results show a clear convergence pattern. By the 70th iteration, the validation accuracies of most clients stabilize, with the majority achieving accuracy levels around or above 0.9. This indicates that our framework successfully synchronizes the learning trajectories across the distributed clients, leading to a consistent and robust overall performance. The increasing mean accuracy, as shown by the trend line, underscores our framework's ability to harmonize client models, even in a decentralized setting.
Despite the overall positive trend, minor fluctuations in accuracy persist for some clients, particularly in the later stages of training. These fluctuations likely result from the heterogeneous nature of the data across clients or the sensitivity of individual models to specific features. While these variations do not detract significantly from the overall convergence, they highlight the necessity of continuous monitoring and potential fine-tuning to ensure that all clients achieve optimal performance.

\subsection{Quantum Noise Analysis}
\begin{figure}[htbp]
    \centering
            \begin{tikzpicture}
            \node[anchor=south west, inner sep=0] (image) at (0,0) {\includegraphics[width=1\linewidth]{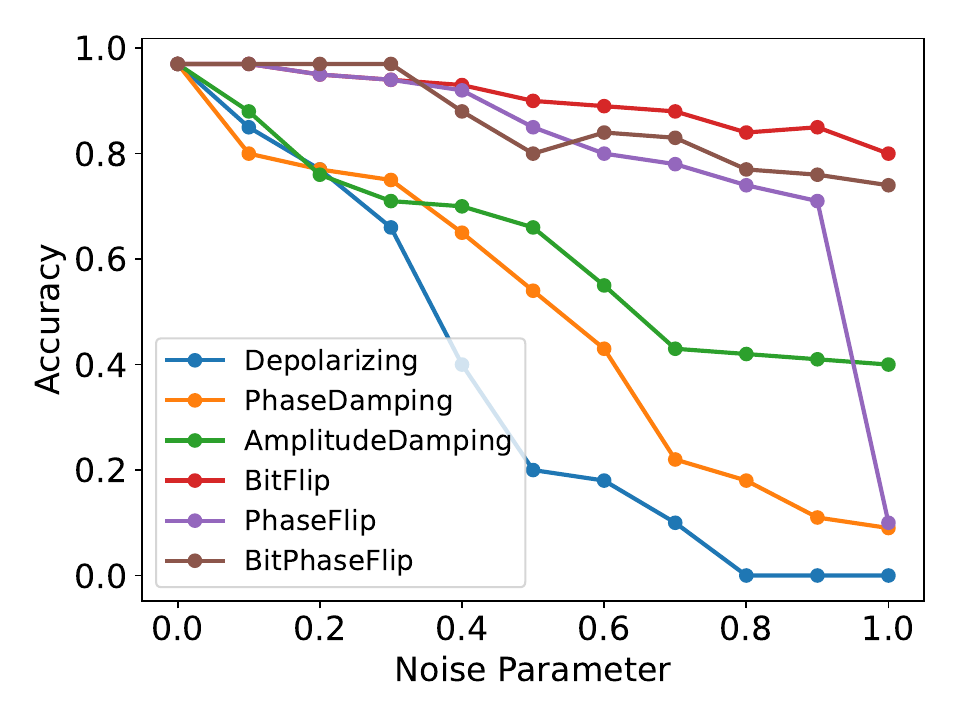}};
            \begin{scope}[x={(image.south east)},y={(image.north west)}]
                \node[circle, draw=black, fill=white, inner sep=1pt] (pointerB) at (0.75,0.68) {\textcolor{black}{3}};
                \draw[-latex, black, thick] (pointerB) -- +(-0.04, 0.09);
                \draw[-latex, black, thick] (pointerB) -- +(-0.04, 0.13);
 \node[circle, draw=black, fill=white, inner sep=1pt] (pointerB) at (0.75,0.43) {\textcolor{black}{2}};
                \draw[-latex, black, thick] (pointerB) -- +(0, 0.08);
               \node[circle, draw=black, fill=white, inner sep=1pt] (pointerB) at (0.44,0.8) {\textcolor{black}{1}};
                \draw[-latex, black, thick] (pointerB) -- +(0.04, 0.09);  
            \end{scope}
        \end{tikzpicture}
    \caption{ 
    Comparative impact of quantum noise models on QFNN-FFD framework accuracy. The graph systematically evaluates the framework's accuracy under the influence of six quantum noise models: depolarizing, phase damping, amplitude damping, bitflip, phaseflip, and bitphaseflip. The noise parameters are adjusted from 0 (indicating no noise) to 1 (signifying maximum noise interference), providing insights into the relative performance stability of the QFNN-FFD framework across a spectrum of quantum noise intensities.}
    \label{noise}
    \vspace{-10pt}
\end{figure}
In our experiments, we expose the QFNN-FFD framework to a spectrum of quantum noise models \cite{noise}, aiming to simulate the challenging conditions of near-term QC devices. As presented in Fig.~\ref{noise}, under the depolarizing noise model, accuracy remains high at 0.97 but plummets to 0 when the noise parameter reaches 1, indicating the model's noise tolerance limit. In the bitflip noise, the  QFNN-FFD shows resilience, maintaining a 0.97 accuracy until the noise parameter hits 0.3 at \textcircled{\raisebox{-0.9pt}1}, after which it drops to 0.8 at a noise level of 1, marking the model's performance threshold. This illustrates how bitflip errors, which flip the state of a qubit, begin to degrade the system's performance only at higher noise levels, demonstrating a strong error tolerance up to a critical point. The amplitude damping noise leads to a less severe decrease in accuracy, from 0.97 to 0.4 at \textcircled{\raisebox{-0.9pt}2} as noise increases, while phase damping impacts it more, reducing accuracy to 0.09, highlighting sensitivity to phase perturbations. These results underscore the QFNN-FFD's varying sensitivity to different types of quantum noise, with phase damping proving particularly detrimental. This sensitivity is crucial for understanding which quantum error correction techniques might be most effective in enhancing the robustness of the model. Remarkably, against phaseflip and bitphaseflip noises, the QFNN-FFD maintains over 0.9 accuracy up to a noise parameter of 0.7 at \textcircled{\raisebox{-0.9pt}3}, only dropping to 0.78, demonstrating significant robustness and potential compatibility with existing quantum technologies. This resilience against phaseflip and bitphaseflip noises suggests that the model's quantum circuit may be naturally more protected against these types of errors, possibly due to the nature of the quantum gates used or the initial state preparation.

Such robustness implies the QFNN-FFD's potential compatibility with current quantum technology, where such noise is prevalent.
The robust performance of the QFNN-FFD across these diverse noise profiles strongly indicates its applicability in quantum-enhanced fraud detection systems. The data clearly illustrates how the QFNN-FFD could provide reliable performance, guiding future enhancements in quantum error correction to fortify the model against the most vulnerable types of noise. These findings are pivotal, as they demonstrate the framework's current efficacy and its potential for adaptation and improvement with the maturation of quantum technologies.
\subsection{Comparison with Existing Works}
\begin{table}[htbp]
    \centering
    \caption{Comparison of QML frameworks on financial fraud datasets.}
\begin{adjustbox}{max width=\linewidth}
\begin{tabular}{llllll}
        \toprule
        Reference & Precision & Recall & F1-Score & Accuracy & Privacy \\
        \midrule
        \cite{ref1} & 84 & 84.44 & 75.68 & 83.92 & $\times$\\
        \cite{ref2} & 96.1 & 79.5 & 86 & 94.5 & $\times$\\
        \cite{kyriienko2022unsupervised} & 90 & -- & -- & -- & $\times$\\
        QNN & 93 & 94 & 94 & 93 & $\times$\\
        \textbf{Our QFNN-FFD} & \textbf{95} & \textbf{96} & \textbf{95} & \textbf{95} & \textbf{\checkmark}\\
        \bottomrule
    \end{tabular}
    \end{adjustbox}
    \label{tab}
\end{table}
The datasets used in the compared studies have been carefully selected using the same attributes as those employed to train our QFNN-FFD, transaction volume, diversity in the variation of values for transactions and merchant categories, and the incidence rates of fraud cases. These are the main selection criteria that ensure uniformity in the level of complexity of datasets and applicability to the particular challenges of financial fraud detection. This careful selection ensures that our comparative analysis is appropriately contextualized and reflects real-world transactional environments and their complexities.

Our purpose is not to compare classical versus quantum methods but specifically to evaluate the performance of our framework against existing QML frameworks for fraud detection. Compared to the results in Table \ref{tab}, our QFNN-FFD outperforms other QML models applied to similar datasets, achieving superior performance metrics. These metrics include precision, recall, F1-score, and accuracy, where QFNN-FFD demonstrates comprehensive superiority across all fronts. This performance is a testament to the model's efficacy and highlights its ability to effectively integrate complex quantum computations within an FL framework.
Unlike the existing models \cite{ref1, ref2, kyriienko2022unsupervised}, which focus solely on performance, QFNN-FFD additionally integrates a privacy-preserving FL approach. This ensures high detection accuracy of 95\% and enhanced data privacy, establishing QFNN-FFD as a leading solution for secure and efficient fraud detection in fintech.
\subsection{Ablation Study}
To further analyze the effectiveness of the proposed QFNN-FFD, we conduct an ablation study comparing its performance against a standalone QNN to evaluate the impact of federated learning on accuracy, privacy preservation, and stability. Both models utilize identical architectures, and hyperparameters, with QNN trained centrally while QFNN-FFD operates across 15 federated clients. The results indicate that QFNN-FFD achieves an accuracy of 95\% compared to 93\% for QNN (see Table \ref{tab}), demonstrating that federated training does not compromise performance while enhancing privacy. Furthermore, QFNN-FFD stabilizes learning across clients, as seen in client-wise validation accuracy trends, mitigating inconsistencies observed in the standalone QNN. Unlike QNN, which requires centralized data aggregation, QFNN-FFD preserves privacy by ensuring data remains local to clients, achieving comparable or superior performance while addressing real-world privacy concerns. These findings validate that QFNN-FFD maintains strong fraud detection capabilities while integrating privacy-preserving mechanisms, making it a promising approach for secure financial applications.
\subsection{Discussion}
Our results show that the framework achieves high validation accuracy, maintains low loss across various operational conditions, effectively harmonizes distributed learning tasks across clients, and exhibits resilience against diverse quantum noise models. Such robustness underlines the framework's suitability for real-world QC environments known for their integral noise issues.
In direct comparison with existing quantum and classical models, QFNN-FFD surpasses typical performance metrics, making it a superior choice for fraud detection. This performance is particularly notable given the framework's integration of privacy-preserving FL, which safeguards sensitive financial data during detection. This dual benefit of enhanced accuracy and increased data privacy sets QFNN-FFD apart as a leading solution for secure and effective fraud detection in the fintech industry.
Furthermore, the framework's ability to maintain high performance under various noise conditions suggests its potential for broader applications beyond financial services, including sectors where data sensitivity and security are paramount. Integrating advanced quantum computational capabilities with robust privacy features positions QFNN-FFD as a scalable solution for future challenges in secure data processing and analysis.
\section{Conclusion}
Our research successfully demonstrates the potential of QFNN-FFD in enhancing fraud detection within the financial sector. By integrating advanced QC techniques with FL, we present a novel approach that significantly improves accuracy and efficiency compared to conventional methods. Our findings reveal that the QFNN-FFD framework, supported by a robust computational infrastructure and optimized through sophisticated preprocessing techniques, can effectively identify fraudulent transactions with high precision.
Its resilience against various quantum noise models is particularly noteworthy, indicating its suitability for real-world application in the near-term QC landscape. This resilience, coupled with the model's ability to maintain high performance under different noise conditions, underscores the practical value of our approach. Furthermore, the QFNN-FFD's adaptability to quantum noise suggests a promising direction for future research in quantum error correction and noise mitigation strategies.
Our study contributes to the emerging field of QC by providing an efficient framework for applying QML while ensuring privacy to solve complex problems in finance. Expanding beyond finance, this framework has the potential to revolutionize fields such as healthcare and cybersecurity, where privacy and data sensitivity are paramount, thus marking a significant milestone in the interdisciplinary application of QML. 
In conclusion, the QFNN-FFD framework addresses key challenges in the fintech sector and also sets a precedent for the deployment of quantum technologies in privacy-critical applications, offering substantial implications for both academic research and industry practices. It encourages further exploration and development within the QC, QML, and FL communities, aiming to unlock new possibilities for handling complex, large-scale data analysis tasks in an increasingly digital and interconnected world.

\section*{Acknowledgments}
 This work was supported in part by the NYUAD Center for Quantum and Topological Systems (CQTS), funded by Tamkeen under the NYUAD Research Institute grant CG008, and the Center for Cyber Security (CCS), funded by Tamkeen under the NYUAD Research Institute Award G1104.
\bibliographystyle{IEEEtran}
\bibliography{reference}

\end{document}